\documentclass[10pt,twocolumn,letterpaper]{article}

\usepackage[top=0.7in,bottom=0.7in,left=0.6in,right=0.7in,columnsep=0.35in]{geometry}

\usepackage{microtype}
\usepackage{graphicx}
\usepackage{subfigure}
\usepackage{booktabs}

\usepackage{hyperref}
\usepackage{hhline}
\usepackage{amsmath,amsthm,amssymb}

\usepackage[utf8]{inputenc}

\title{Applying Bayesian Hierarchical Probit Model to Interview Grade Evaluation}

\author{Yuki Ohnishi$^1$ and Shinsuke Sugaya$^2$}
\date{%
    $^1$Purdue University and%
    $^2$Bizreach, inc.\\[2ex]%
}
\begin{document}

\maketitle

\begin{abstract}
Job interviews are a fundamental activity for most corporations to acquire potential candidates, and for job seekers to get well-rewarded and fulfilling career opportunities. In many cases, interviews are conducted in multiple processes such as telephone interviews and several face-to-face interviews. At each stage, candidates are evaluated in various aspects. Among them, grade evaluation, such as a rating on a 1-4 scale, might be used as a reasonable method to evaluate candidates. However, because each evaluation is based on a subjective judgment of interviewers, the aggregated evaluations can be biased because the difference in toughness of interviewers is not examined. Additionally, it is noteworthy that the toughness of interviewers might vary depending on the interview round. As described herein, we propose an analytical framework of simultaneous estimation for both the true potential of candidates and toughness of interviewers' judgment considering job interview rounds, with algorithms to extract unseen knowledge of the true potential of candidates and toughness of interviewers as latent variables through analyzing grade data of job interviews. We apply a Bayesian Hierarchical Ordered Probit Model to the grade data from HRMOS, a cloud-based Applicant Tracking System (ATS) operated by BizReach, Inc., an IT start-up particularly addressing human-resource needs in Japan. Our model successfully quantifies the candidate potential and the interviewers' toughness. An interpretation and applications of the model are given along with a discussion of its place within hiring processes in real-world settings. The parameters are estimated by Markov Chain Monte Carlo (MCMC). A discussion of uncertainty, which is given by the posterior distribution of the parameters, is also provided along with the analysis.
\end{abstract}

\maketitle

\section{Introduction}
The analyses described in this report address challenges related to estimating the true potential of candidates and the toughness of interviewers during job interviews. Job interviews play a key role in most companies for selection of candidates, and for job seekers to find appropriate career opportunities. Although important for human resource management, it is often the case in most business fields that interviews are conducted based on interviewers' gut feelings. In such cases, feedback can be biased because each interviewer has an idiosyncratic standard for evaluation: often they do not agree among themselves on what questions should be asked and how applicants are evaluated. For companies to make better use of feedback and to make more reasonable decisions in hiring promising candidates, more robust and reliable methodologies must be used to grade candidate potential. Nevertheless, when considered from the perspective of a job seeker, they should be evaluated with a unified framework that can estimate their true potential, which would keep them from missing possible career opportunities. Although a candidate's potential and an interviewer's toughness are very important factors for interview evaluations, it is not easy to estimate them because each interviewer interviews different candidates. Aggregating grade data across interviewers systematically can understate the ability of an interviewer rating to predict candidate performance. A more sophisticated statistical method must be used to estimate them simultaneously.

After some screening processes by resume, job interviews generally take place in multiple rounds, such as casual coffee meetings, telephone interviews and face-to-face interviews. The interviewers then grade candidates on each round. It is noteworthy that the toughness of interviewers' judgment may vary among rounds. Intuitively, latter interviewers may try to give thorough evaluations, which can engender more negative feedback. Therefore, one must consider the variance of each round when estimating interviewers' toughness.

As described above, standardization of job interview grades and hiring decisions is an important problem for any moderately sized organization. However, we often rely on expert judgments to ascertain whether to hire a certain candidate. These difficulties must be addressed in a quantitative manner to assign the right interviewers to the right candidates for the right rounds.

It is noteworthy that an ideal evaluation framework achieves the following goals simultaneously: (1) estimating candidates' true potential and interviewers' toughness simultaneously from grading data, with consideration of mutual interaction; (2) estimating the variance of an interviewer's judgment according to job interview rounds; (3) predicting the distribution of interviewer's judgment about candidates and automatically assigning the right interviewer at the right time. The first two are attempts at standardizing job interview grading, while the third particularly addresses the application of this framework.

In order to achieve these goals, we applied a Bayesian Hierarchical Ordered Probit Model (BHOPM) to real-world data. This method itself has little novelty, but the focus of this paper, and our contributions are on the applied side. We demonstrate how naturally BHOPM can accommodate our problem settings. The contributions of this paper are as follows.

\begin{itemize}
  \item A new application area to the rich literature of Bayesian Hierarchical Ordered Probit Model to evaluate latent candidate potentials and interviewers' judgment toughness in consideration of the variance of interview's rounds.
  \item Extensive evaluations of the proposed approach. The result showed the model successfully estimated the toughness of interviewers.
  \item Interpretations of experimentally obtained results with consideration of model uncertainty by Bayesian inference. Examples are shown about how our model can help us make decisions on hiring.
\end{itemize}

As described herein, we propose an analytical framework of simultaneously estimating latent candidate potential and interviewers' judgment toughness in consideration of the variance of interview's rounds, through analysis of ordinal grade data of job interviews. The reminder of this paper is structured as follows. Section 2 describes our problem setting and our model. Section 3 presents experiments with real-world data and experimental results. Section 4 presents discussions of interpretation, giving responses to several research questions with the application samples. We surmise that many of these answers can be generalized beyond our scenarios. Section 5 explains several works related to this study. Section 6 concludes this paper.

\section{Problem Setting and Methodology}
This section provides details of our problem setting in interview evaluation, and our approach to solve that problem. As described herein, Bayesian Hierarchical Ordered Probit Model (BHOPM) is used to analyze grading data on a scale from 1 (Failure) to 4 (Excellent). We present a mathematical formulation of our model and how it naturally accommodates our problem.

As described in Section 1, this study has two main purposes. The first is to contrive an analytical framework of standardizing job interviews and its evaluation processes. Many companies evaluate job applicants using a grade score. Although this scoring approach is simple and easy to understand for most people, a problem exists by which the grade is fundamentally dependent on each interviewer. As stated already, the purpose of standardizing job interviews and evaluation processes is to evaluate candidates' true potential without interviewer bias or interview round bias. To achieve this goal, we formulate this problem of estimating some latent variables such as a candidate's potential and an interviewer's bias: the interviewer's toughness. A detailed formulation will be discussed in the next subsection.

The estimated latent variables of candidates and interviewers, with consideration of generating process of grade data, have widely diverse applicability. Presentation of some applications of these variables and how they are interpreted is also included in the scope of this study, which is shown in Section 4.

\subsection{Baysian Hierarchical Ordered Probit Model}
Estimating a candidate's potential and an interviewer's toughness from interview grade data can be naturally modeled with a Bayesian Hierarchical Ordered Probit Model (BHOPM), which is an extended model of an Ordered Probit Model with hierarchical structures. It can naturally accommodate differences in individual judgments, and be inferred in a Bayesian manner.

Presuming $K$ ordinal categorical values, an Ordered Probit Model presents a mode of predicting ordinal categorical measure $k (k=1,\ldots,K)$ based on other predictors $x$. To accomplish this, one can introduce linear regression on x and then link the continuous prediction to ordinal value $k$ via a thresholded cumulative-normal function. Figure 1 portrays the underlying mappings from $x$ to $k$ [doing Bayesian data analysis]. As shown in Figure 1, the predictor value, $x$, becomes mapped to an underlying metric with a linear relation, $\mu= \beta_0+ \beta x$. The underlying metric value is noisy, and is distributed around $\mu$ as a normal distribution with standard deviation $\sigma$. Finally, the underlying metric scale is carved into intervals by several thresholds, which are also estimated from the data. When the underlying metric value falls between thresholds $\theta_{k-1}$ and $\theta_{k}$, an ordinal value $k$ is generated. We denote $\Phi$ as a normal cumulative function. Then, the mass of the normal distribution between $\theta_{(k-1)}$ and $\theta_{(k)}$ is calculated as $\Phi ((\theta_{k} - \mu) / \sigma) - \Phi ((\theta_{k-1} - \mu) / \sigma)$. This mass can be interpreted as the probability of generating an ordinal value $k$. For the lowest and highest ordinal values, the outside thresholds are negative and positive infinity. Therefore, the probability is $\Phi ((\theta_{1} - \mu) / \sigma)$ for the lowest and 1-$\Phi ((\theta_{K-1} - \mu) / \sigma)$ for the highest. Additional information related to Ordinal Probit Models is available in the literature \cite{Kruschke14}.
\begin{figure}[h]
  \centering
  \includegraphics[width=\linewidth]{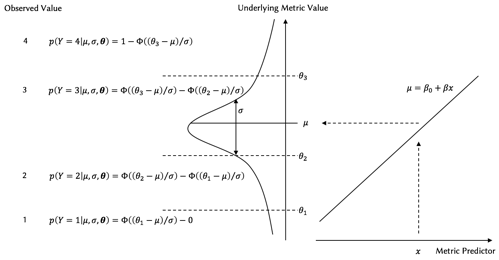}
  \caption{Overview of underlying mapping in ordinal probit regression.}
\end{figure}

Next, we consider how our observed grade data is generated. When it comes to evaluating candidates, we should mentally hold some continuous values that represent the candidate's evaluation, which is affected by the candidate's potential and the interviewer's bias. Then we try to map the continuous rating value $\mu_{n}$ to ordinal categorical grade metrics, with individual thresholds determining the grade to which the value belongs. This generative process can be modeled quite naturally by BHOPM. In this context, we consider a linear combination of latent variables as
\begin{equation} \label{eq:1}
  \mu_{n} = \mu_0+\mu_{c}+\mu_{i}
\end{equation}
where $\mu_0$, $\mu_{c}$ and $\mu_{i}$ respectively represent constant term, latent variables for candidate performance, and the interviewer's bias at interview. The underlying metric value is distributed around $\mu_{n}$ as a normal distribution with standard deviation $\sigma$, which we will explain hereinafter.

Furthermore, as already stated, candidate performance and interviewer's toughness may vary depending on interview rounds. To express this circumstance, we presume that these latent variables are generated as presented below.
\begin{equation} \label{eq:2}
  \mu_{c} \sim Normal(\alpha_{c}+\gamma_{c,r},\sigma_{c})
\end{equation}
\begin{equation} \label{eq:3}
  \mu_{i} \sim Normal(\beta_{i}+\delta_{i,r},\sigma_{i})
\end{equation}
\begin{equation} \label{eq:4}
  \alpha_{c} \sim Normal(0,\sigma_{\alpha})
\end{equation}
\begin{equation} \label{eq:5}
  \beta_{i} \sim Normal(0,\sigma_{\beta})
\end{equation}
\begin{equation} \label{eq:6}
  \gamma_{c,r} \sim Normal(0,\sigma_{\gamma})
\end{equation}
\begin{equation} \label{eq:7}
  \delta_{i,r} \sim Normal(0,\sigma_{\delta})
\end{equation}
Therein, we define $\alpha_{c}$ $(c=1\ldots C)$ and $\beta_{i}$ $(i=1\ldots I)$ as candidate potential and interviewer toughness; we also assume that they follow a normal distribution respectively. Parameters $\gamma_{c,r}$ and $\delta_{i,r}$ $(r=1\ldots R)$  are defined as an adjustment on each interview round. Equation \ref{eq:2} defines that $\mu_{c}$ comprises of $\alpha_{c}$ and $\gamma_{c,r}$ and is distributed around their summation with variance $\sigma_{c}$. The same definition applies to $\mu_{i}$. 

Equation \ref{eq:4}, \ref{eq:5}, \ref{eq:6} and \ref{eq:7} demonstrate that these parameters have hierarchical structures. The parameters for individuals are generated from a common popular distribution. In our problem setting, it might be reasonable to expect that estimates of $\alpha_{c}$ are mutually related because we can assume that individual parameters are, to some extent, mutually close. The same is true of the other parameters, $\beta_{i}$, $\gamma_{c,r}$ and $\delta_{i,r}$. We know that the interviewers review the candidate in various aspects depending on interview rounds but we do not expect that they are completely independent. To express this sort of relation, we posit that these parameters have a hierarchical structure by which each parameter for individuals is viewed as a sample from a common popular distribution, as described in Equation \ref{eq:4}, \ref{eq:5}, \ref{eq:6} and \ref{eq:7}. We set a normal prior to each parameter here and assume a Cauchy prior for each deviation for robustness \cite{Gelman13}, \cite{Gelman08}, as shown below.
\begin{equation} \label{eq:8}
  \sigma_{c} \sim Cauchy(0, 2.5)
\end{equation}
\begin{equation} \label{eq:9}
  \sigma_{i} \sim Cauchy(0, 2.5)
\end{equation}
\begin{equation} \label{eq:10}
  \sigma_{\alpha} \sim Cauchy(0, 2.5)
\end{equation}
\begin{equation} \label{eq:11}
  \sigma_{\beta} \sim Cauchy(0, 2.5)
\end{equation}
\begin{equation} \label{eq:12}
  \sigma_{\gamma} \sim Cauchy(0, 2.5)
\end{equation}
\begin{equation} \label{eq:13}
  \sigma_{\delta} \sim Cauchy(0, 2.5)
\end{equation}

We conducted some experiments to fix the hyperparameters for the Cauchy distributions. We set the variance to 2.5 because it achieved the lowest WAIC \cite{Watanabe13_1}, \cite{Watanabe13_2} and all parameters successfully converged. We will have a detailed discussion on the convergence in Section 4. We assume that choosing $2.5$ as the variance for the Cauchy distribution is reasonable because it is large enough to cover the estimators of our interest. Once a continuous rating value towards a candidate is generated using the process explained above, we try to map the continuous value to categorical grade values with each threshold. Even if some interviewers have almost identical continuous rating values towards a certain candidate, some generous ones might give $3$, whereas the others give $2$. Consequently, it might be reasonable to posit that each interviewer has a personal threshold. To model this, the threshold of each interviewer is assumed as a sample from uniform prior distribution defined as
\begin{equation} \label{eq:14}
  \theta_{i,k} \sim Uniform(\theta_{i,k-1}, \theta_{i,k+1})
\end{equation}
where $\theta_{i,k}$ represents the threshold of interviewer $i$, which categorizes the grade into ordinal value $k$ or $k+1$. When the continuous rating value falls at the left side of $\theta_{i,k}$, ordinal value $k$ is generated. 

Thresholds $\theta_{i,k}$ and a cumulative normal function of underlying metric value $\mu_n$, with deviation $\sigma$, define the probability of generating each categorical output. This probability can be interpreted as the parameter $\pi_{n,k}$ of a categorical distribution, from which we can assume that an observed grade is sampled. It can be formulated as follows.
\begin{displaymath}
  \pi_{n,k}=\Phi ((\theta_{i,k} - \mu_n) / \sigma) - \Phi ((\theta_{i,k-1} - \mu_n) / \sigma) \quad(k=2,\ldots,K-1)
\end{displaymath}
\begin{displaymath}
  \pi_{n,1}=\Phi ((-1 - \mu_n) / \sigma)  \quad(\because \theta_{i,1}= -1) 
\end{displaymath}
\begin{equation} \label{eq:15}
  \pi_{n,K}=1-\Phi ((1 - \mu_n) / \sigma)  \quad(\because \theta_{i,K}= 1) 
\end{equation}
\begin{equation} \label{eq:16}
  \sigma \sim Cauchy(0, 2.5)
\end{equation}
\begin{equation} \label{eq:17}
  y_n \sim Categorical(\pi_{n,k})
\end{equation}
In those equations, $y_n$ is an observed grade data on a scale from $1$ to $4$ in our case. We have four categories of $y_n$. Therefore, we require three thresholds to define parameters $\pi_{n,k}$ for each category. We fix $\theta_{i,1}=-1$ and $\theta_{i,K}=1$ for the identification of the model's parameters \cite{Kruschke14}. 

\begin{figure}[h]
  \centering
  \includegraphics[width=\linewidth]{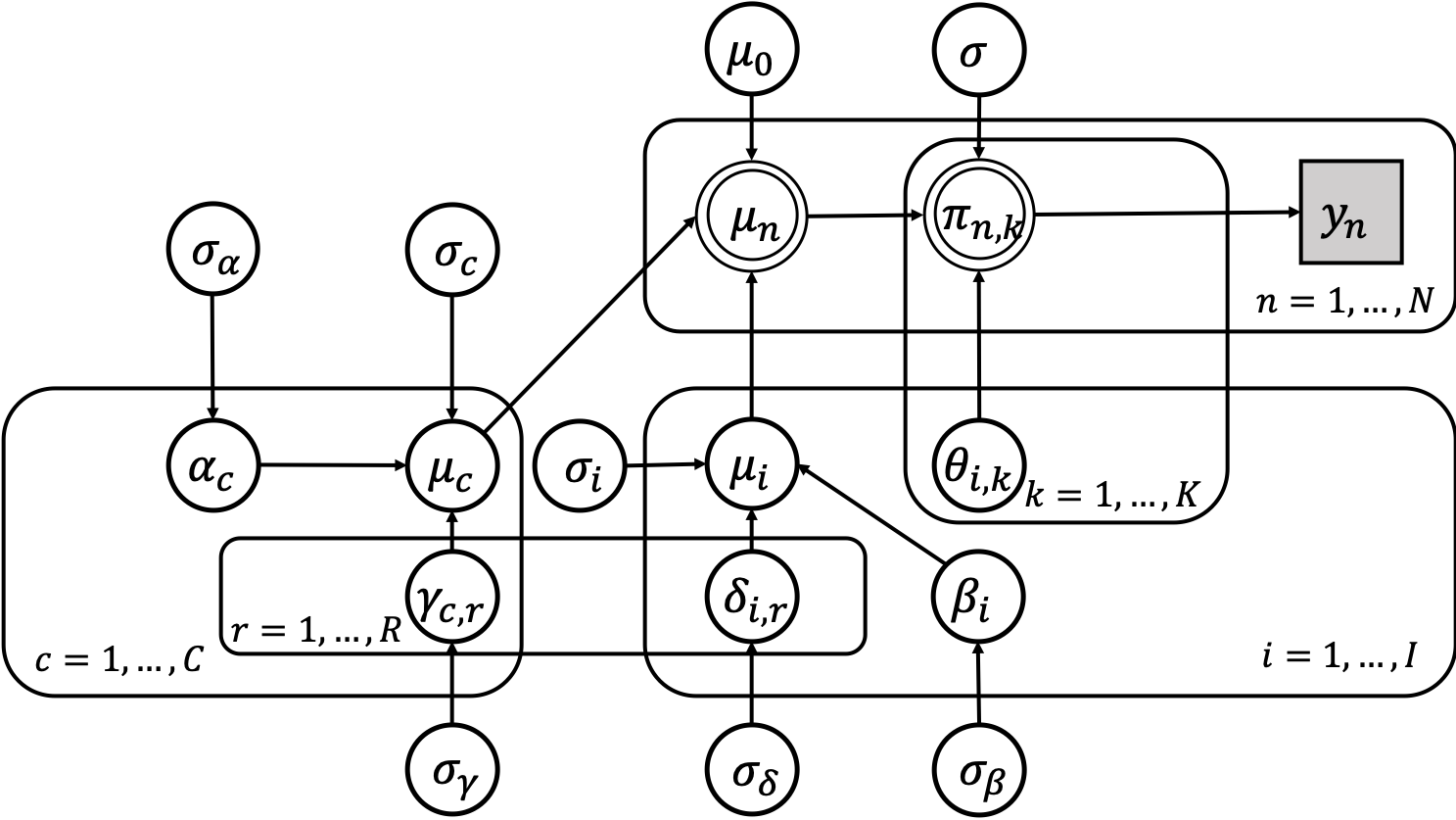}
  \caption{Graphical model.}
  \label{fig:graphmodel}
\end{figure}

The graphical model for the model we described in this section is given in Figure \ref{fig:graphmodel}.

\subsection{Inference}

We estimate all estimands using Bayesian inference. The previous subsection introduced our model. We inferred all parameters by Markov Chain Monte Carlo (MCMC) sampling with Stan \cite{Carpenter}. Stan, a state-of-the-art platform that is useful for statistical modeling and high-performance statistical computation, offers a probabilistic programming language in which users specify log density functions of models and conduct full Bayesian statistical inference with MCMC sampling. The default internal algorithm is N-U-Turn Sampler (NUTS) \cite{Hoffman}, an extension of Hamiltonian Monte Carlo (HMC) \cite{Neal}, which eliminates the need to set the problematic number-of-steps parameter. Bayesian inference is widely used to estimate parameters of complicated models, for which the posterior distributions cannot be solved analytically. One can estimate a Bayesian credible interval on the parameters by sampling from the posterior distribution over the parameters, which enables us to assess model uncertainty.

\section{Experiments}

This section presents verification of the proposed model capacity for real-world data generated at HRMOS; an Applicant Tracking System (ATS) operated by BizReach, Inc. We first provide details of the dataset we collected for experiments.

\subsection{HRMOS and Data Collection}
HRMOS is an integrated system that has several submodules for human resource management. Among these, HRMOS has a submodule for applicant tracking, which enables corporate recruiters to manage the whole process of job interviews seamlessly: from interviewer assignment to candidate evaluation.

We collected about 10,000 interview grade data from January 1, 2017 through December 27, 2017 at BizReach, Inc. Interviewers graded candidates with a scaled score from 1 to 4. Each datum has a flag that shows the interview round, such as a meeting, interview or final interview. A summary of collected data is presented in Table \ref{tab:1}. 

\begin{table}
  \caption{Summary of Collected Data}
  \label{tab:1}
  \begin{center}
  \begin{tabular}{|l|l|}
    \hline
    \#Total Interviews           & 10095 \\ \hline
    \#Unique Candidates          & 3046  \\ \hline
    \#Unique Interviewers        & 202   \\ \hline
    \multicolumn{2}{|l|}{Round Summary}  \\ \hline
    \#Meetings (Round 1)         & 5594  \\ \hline
    \#Interviews (Round 2)       & 3846  \\ \hline
    \#Final Interviews (Round 3) & 655   \\ \hline
    \multicolumn{2}{|l|}{Grade Summary}  \\ \hline
    \#Grade 1 (Failure)          & 1926  \\ \hline
    \#Grade 2 (Not Bad)          & 5199  \\ \hline
    \#Grade 3 (Good)             & 2845  \\ \hline
    \#Grade 4 (Excellent)        & 125   \\ \hline
\end{tabular}
\end{center}
\end{table}

\section{ANALYSIS AND APPLICATIONS}

This section presents extensive analyses and some applications in real-world business settings. First, we explain the analyses and interpretations of the posteriors of estimated parameters described in Section 3. Additionally, we present widely diverse cases for applicability of our model. We demonstrate how to predict unseen evaluations and interpretation of how much the candidate potential will change after interview.

As stated already in the previous section, the whole parameter inference was conducted by MCMC with Stan. Several MCMC parameters must be fixed before running a simulation. These parameters are presented in Table \ref{tab:2}. We used ver. 2.16.0 of Stan with the Python interface: PyStan \cite{Carpenter}.

\begin{table}[]
\caption{Simulation Parameters}
  \label{tab:2}
  \begin{center}
\begin{tabular}{cc}
\hline
\textbf{Number of chains}  & 4    \\
\textbf{Burn-in}           & 1000 \\
\textbf{Number of samples} & 4000 \\
\textbf{MCMC algorithm}    & NUTS
\end{tabular}
\end{center}
\end{table}

\subsection{Model Evaluations}

The main purpose of our model is to estimate candidate's true potential and interviewer's toughness as latent variables. Before going into further analyses on these values, we should validate our model. For validation, we should evaluate the interviewer's estimand rather than the predictive power of the model due to our focus on its interpretability. Interview's toughness $\beta_{i}$ is evaluated by comparison to a toughness ranking judged by 12 members in the human resource team of our company. Each member is randomly assigned 30 interviewers to evaluate, and then asked to rank them in terms of toughness. We calculated Spearman's rank correlation coefficient between maximum a posteriori (MAP) estimates of each interviewer's toughness $\beta_{i}$ and the rankings members gave. The maximum, median and minimum coefficients are calculated as 0.72, 0.52 and 0.30, respectively. The interviewer toughness is shown to correlate with human judgment about interviewers, in comparison with the inferred toughness. We can conclude that the estimates are the good representation of the true toughness.

In order to evaluate the convergence of MCMC sampling, we calculated the Gelman-Rubin statistic $\hat{R}$. $\hat{R}$ is the potential scale reduction factor on split chains (at convergence, $\hat{R}=1$) \cite{Gelman13}. In general, $\hat{R} < 1.1$ is a good indicator of convergence of the parameter estimation. The proposed model has a hierarchical structure with thousands of parameters for all candidates and interviewers. We confirmed that all parameters in our model have $\hat{R}$ less than or equal to $1.02$. We can conclude that all parameters for our model successfully converged. The log likelihood also converged with $\hat{R}=1.08$.

Finally, we evaluated the model's predictive performance using a train-test split scheme. As test data, we collected 387 interview grade data from January 1, 2018 through January 31, 2018. The data size is limited compared to the train data because with time series data, particular care must be taken in splitting the data in order to prevent data leakage. We must withhold all data about events that occur chronologically after the events used for fitting the model. We fit our model with the training data shown in Table \ref{tab:1} and validate the predictive performance of the model using the test data. This predictive problem can be regarded as a multi-label classification problem. In this problem setting, we can estimate a predictive distribution of the grade value in Bayesian inference conditioning on observed grade data $y$, Monte Carlo inference, as shown below.

\begin{equation} \label{eq:pred}
  p(y^*|y)=\int p(y^*|\theta, y)p(\theta|y)d\theta
\end{equation}

\begin{figure}[h]
  \centering
  \includegraphics[width=\linewidth]{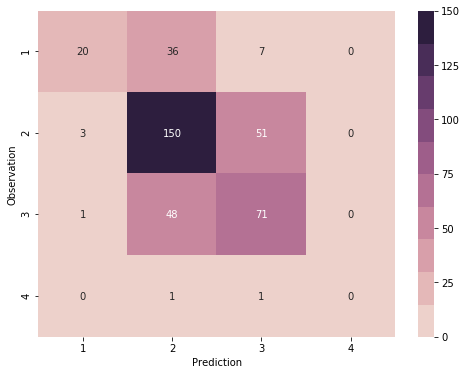}
  \caption{Confusion matrix.}
  \label{fig:confmat}
\end{figure}

In this equation, $p(y^*|y)$ represents the predictive distribution of $y^*$, the predicted grade for test data, conditioning on $y$, observed training grade data, and $\theta$ denote all parameters of our model, as given from Equation \ref{eq:1} to \ref{eq:17}. All parameters $\theta$ are marginalized using Monte Carlo sampling. We sampled 4,000 values from the joint posterior \ref{eq:pred} and calculated the probability of the data categorized in each label. Then, we designated the category with the highest probability as the predictive value. Figure \ref{fig:confmat} shows the confusion matrix. It indicates that 64\% of the data fall within the correct categories and 93\% of them fall with the correct or the adjacent categories. Although the predictive power is not necessarily the main focus of our work, we can say that the model has a decent predictive power.

\subsection{Posterior Distributions of Candidates and Interviewers}

\begin{figure}[h]
  \centering
  \includegraphics[width=\linewidth]{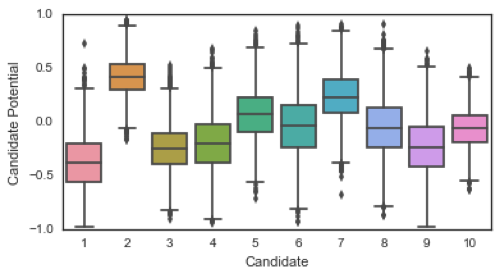}
  \caption{Distributions of candidate potential.}
  \label{fig:fig2}
\end{figure}

As stated already, estimating candidate potential and interviewer bias, which are defined in Equation \ref{eq:4} and \ref{eq:5}, is one purpose of this study. Figure \ref{fig:fig2} presents the posteriors of 10 randomly sampled candidates. Regarding candidate potential, distributions with higher values of $\alpha_{c}$ represent a higher potential of the candidate. This result demonstrates that candidate 2 is more promising than other candidates.

\begin{figure}[h]
  \centering
  \includegraphics[width=\linewidth]{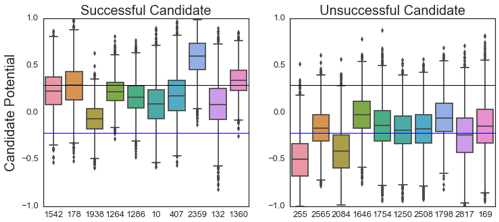}
  \caption{Difference between distributions of successful and unsuccessful candidates.}
  \label{fig:fig3}
\end{figure}

Figure \ref{fig:fig3} presents posteriors of successful and unsuccessful candidates and the mean of their MAP estimates. The upper horizontal axis is the mean of successful candidates, whereas the lower axis is the mean of unsuccessful candidates. As readily apparent, the potential of successful candidates is precisely estimated as higher than that of the unsuccessful candidates. 

\begin{figure}[h]
  \centering
  \includegraphics[width=\linewidth]{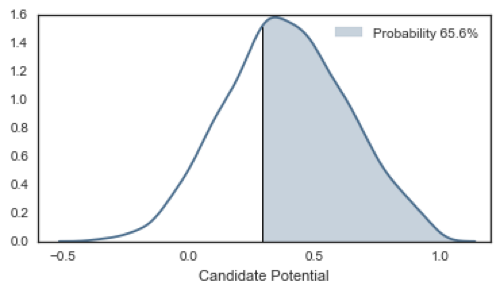}
  \caption{Uncertainty of candidate potential and probability above the defined threshold.}
  \label{fig:fig4}
\end{figure}

This threshold may be useful in determining whether interviewers let the candidate pass the interview or not. That is, one calculates the posterior of the candidate potential after an interview. If the posterior surpassed the successful threshold, then one can allow the candidate to pass the interview. Moreover, as might be readily apparent, all estimands are estimated with a Bayesian credible interval. A long tail distribution reflects that the model is not confident of its estimation because of the small size of the data. One should consider this uncertainty when evaluating them. For example, let us assume a candidate has the posterior distribution shown in Figure \ref{fig:fig4}. In this case, it can be assumed that the potential of the candidate surpasses the successful threshold with 65.6\% confidence, which is shown as the shaded area in Figure \ref{fig:fig4}. This threshold is merely an example in our case. One can decide what is to be used as a threshold according to a business scenario. 

\begin{figure}[h]
  \centering
  \includegraphics[width=\linewidth]{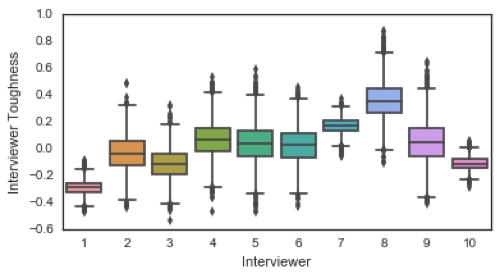}
  \caption{Distributions of interviewer toughness.}
  \label{fig:fig5}
\end{figure}

Figure \ref{fig:fig5} shows the posteriors of randomly sampled interviewers. Regarding interviewers, tougher interviewers have lower values. Therefore, interviewer 1 is the toughest interviewer depicted in this figure. One can infer that the estimations of interviewers 1, 7, and 10 are confident with a sharp distribution compared to others.

\subsection{Posterior Distributions of Latent Variables For Rounds}

\begin{figure}[h]
  \centering
  \includegraphics[width=\linewidth]{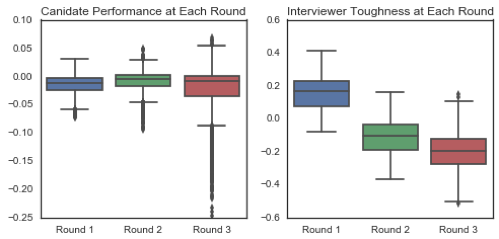}
  \caption{Distributions of round bias.}
  \label{fig:fig6}
\end{figure}

The estimands presented above were estimated without interview round bias. Figure \ref{fig:fig6} shows how biased the candidate performance and interviewer toughness are for each interview round. The candidate performance does not vary considerably between rounds. However, the interviewers become tougher as the interviews approach the end. This result follows our intuition that the interviewers in the latter rounds try to evaluate the candidates in depth, which can engender more negative feedback. The first round has a rather positive impact on the evaluation, which suggests that we should not blindly trust the evaluation in the first round.

\subsection{Predictive Grade for Candidates}
\begin{figure}[h]
  \centering
  \includegraphics[width=\linewidth]{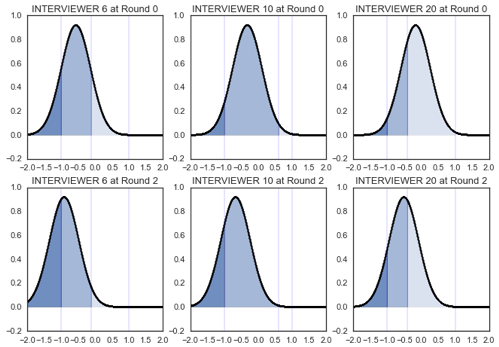}
  \caption{Predictive distributions of grade via point estimates. (The different levels of blue shade show the predicted probability of each grade.)}
  \label{fig:fig7}
\end{figure}

In this subsection and the next subsection, we move deeper into an explanation of more practical aspects with the assumption that all latent variables for candidates and interviewers, described in the previous sections, are already estimated in daily batch processing. A question people might ask is what grade is assigned in future interviews. We presume a hypothetical setting in which we have a new candidate to evaluate. We want to know what grade each interviewer gives. Certainly, we know little about the candidate: no prior distribution is known for the candidate potential. In that case, maybe we should assume an average potential for them. How can one estimate an average candidate potential $\alpha_{ave}$? We calculated it as shown below.

\begin{equation} \label{eq:18}
  \alpha_{ave}=  \sum_{c=1}^{C} \alpha_{c}^{MAP}/C 
\end{equation}

where $C$ denotes the total number of candidates; $\alpha_{c}^{MAP}$ represents a MAP estimate of candidate potential $\alpha_{c}$. To make the point more visually and conceptually understandable, we use point estimates for all latent variables and infer the predictive distribution portrayed in Figure \ref{fig:fig7}. As for the $\alpha_{ave}$ estimate, we merely average all MAP estimates of $\alpha_{c}$. 

In Figure \ref{fig:fig7}, upper panels portray the predictive distributions of some interviewers at Round 0. The lower panels depict those at Round 2. The predictive distribution for each interviewer moves slightly to the left at Round 2. This is the effect of round bias. Additionally, each interviewer has a threshold. The left shaded area equals the probability of grade 1. The right area, which is too slim to be visible, equals the probability of grade 4. In this figure, Interviewer 6 is the toughest in terms of determining failure or not. He assigns grade 1 to an average candidate at Round 2 with about 45\% confidence. Interviewer 10 has a high threshold between grade 2 and grade 3. Therefore, he does not easily assign a better grade than grade 2. Compared to the two, Interviewer 20 gives more generous evaluations.

\subsection{Change of Candidate Potential After Evaluation}

\begin{figure*}[h]
  \centering
  \includegraphics[width=\linewidth]{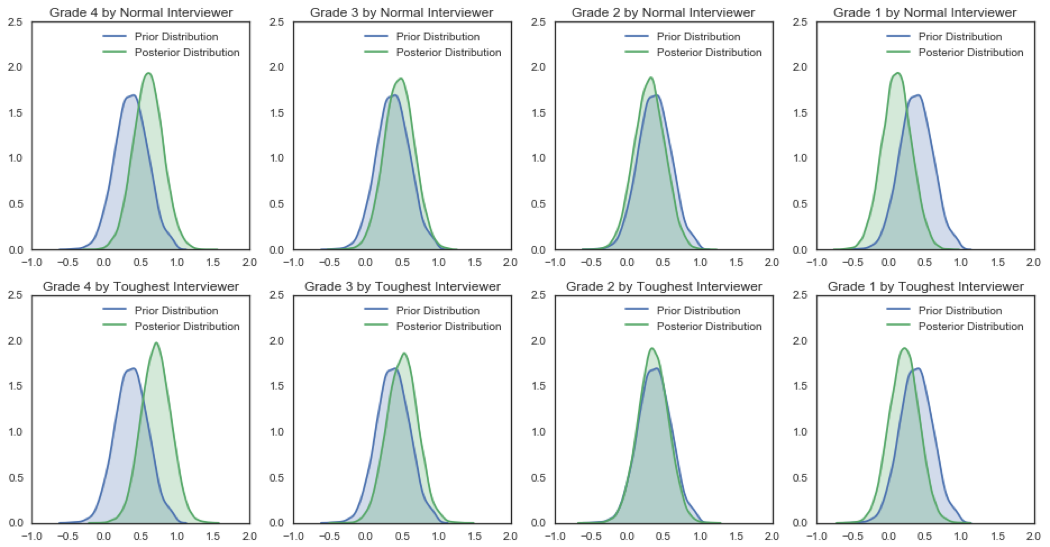}
  \caption{Change of candidate potential distributions after being graded.}
  \label{fig:fig8}
\end{figure*}

Another question one might ask is how much the estimated candidate potential would change if a certain interviewer graded the candidate. Consider the case in which one must assign the final interviewer to a certain candidate, and decide whether to hire the candidate based on the judgment of this interviewer. In such a scenario, one might want to update the candidate potential considering the interviewer's toughness and the candidate potential, each of which has already been estimated. Our model can accommodate this situation naturally by evaluating the posterior distribution of the candidate potential after the interviewer assigns the grade. This functionality might assist the automatic screening by defining thresholds of successful candidates.

We simulate this situation using empirical Bayes method. The distribution of previously estimated parameters, such as potential, toughness, round bias and threshold, are approximated to follow a normal distribution. The parameters of the normal distributions are calculated using maximum likelihood method. After defining the distribution, we use it as the prior distribution of our model's parameters, such as candidate potential, round bias, and threshold; then we simulate how the candidate potential variable might change. Figure \ref{fig:fig8} presents a change in predictive distribution of a random candidate potential after evaluation by a normal interviewer and a tough interviewer. Those figures have the same prior distribution in blue in common and updated posterior distribution in green. The lower left panel displays the posterior when the toughest interviewer grades candidate 4 (Excellent). As might be readily apparent, the distribution moves markedly to the right compared to a normal interviewer's grade 4 feedback, which means that an excellent grade by a tough interviewer can be positive feedback for candidate evaluation. However, the upper right panel shows the posterior after the easiest positive feedback for candidate evaluation.

The upper right panel portrays the posterior after the easiest interviewer grades candidate 1 (Failure), which engenders a further move to left compared to grade 1 by the tough interviewer (lower right). From this result, we infer that bad evaluations from easy interviewers have a more negative impact on candidate evaluation than tough interviewers.

\section{RELATED WORKS}
This study specifically examined the application of Bayesian Hierarchical Ordered Probit Model, which has an extremely wide range of applicability in diverse scientific fields, such as finance, social science, and medical science. In the context of finance, one report \cite{Hausman92} describes estimation of the conditional distribution of trade-to-trade price change using an ordered probit model. They estimate the model via maximum likelihood. In medical science, one report of the relevant literature \cite{Munkin07} presents a Bayesian ordered probit model to analyze the effects of different medical insurance plans on the level of hospital utilization, allowing for potential endogeneity of insurance status. In the field of social science, another study \cite{Rampichini} applied Bayesian hierarchical ordered probit model to analyze life satisfaction in Italy. This study treats survey data with a group structure and an ordinal response variable. They inferred that this group structure, which they define as regional contexts, affects the individual life satisfaction level.

As for studies related to employment interview, one \cite{Judge} summarized the history of this research area. Some studies specifically examine the effects of candidate-related factors during an interview, such as smiling, accents, and speech duration, on interview results \cite{Krumhuber}, \cite{Willemyns}, \cite{Matarazzo}. Another researcher \cite{Dreher} has argued that because interviewers differ in their evaluations and because they use different parts of a rating scale, aggression of ratings across interviewers systematically understates the ability of interviewer ratings to predict job performance. It has been argued that the reliability of job interview is low because interviewers do not agree among themselves in terms of what questions should be asked and how applicants are evaluated.

The main prior work of our study \cite{Pulakos}, is a study of estimation of individual differences in interviewer ratings and standardization of interview process. This study conducted a comprehensive investigation of individual differences in interviewer validity. They analyzed the decisions of 62 interviewers and described correlations between an individual interviewer's ratings and job performance for job interviewees who were hired. Aside from this work, meta-analysis of the relations between individual assessments and job performance was conducted in at least one earlier studies \cite{Morris}. We have the same awareness of this study, but we pursued a different approach than theirs in the sense that our approach specifically examines modeling and its simultaneous simulation of interviewers' and candidates' differences including round bias.

Ordered probit model has a long history of application, but Bayesian hierarchical approaches are an emerging area\cite{GelmanHill}. Related works are few. No existing work specifically examines bias problem in employment interview. As stated in one report of the literature \cite{Judge}, quantitative evaluation of promising candidates and individual differences in interviewer validity is necessary for corporate activity for identifying and evaluating the right people. We have added a new application area to the rich literature of ordered probit model and employment interview research \cite{Greene}. 

Our work is part of a growing body of work advancing the use of statistics in real-world businesses of various fields such as marketing science \cite{yuki2020} and sociology \cite{Muller}, \cite{Letham}, \cite{Green}. These works, and ours, address needs for statistical methodologies in real-world business.

\section{CONCLUSION}
As described in this paper, we demonstrated an analytical framework of estimating true candidate potential and interviewer's toughness at job interview without interview round bias. We formulated this problem as estimating latent variables of a hierarchical Bayesian ordered probit model, and could successfully estimate the candidate potentials and the interview toughness. We also provided the interpretations of these parameters and conducted extensive analyses of these. One of the most interesting findings in this research is the fact that, from various point of views, job interviews are biased. Each interviewer has the different level of toughness and, as we imagined, the interview rounds affect the interview results.  Overcoming these biases might benefit both companies and candidates.

Some useful applications are also presented in a real business setting. This study presents a new approach to analyzing job interview data and proposes some implications for understanding candidates' true potential and interviewer bias. This approach has widely various potential applications in any organization that manages job interview grade data as described herein.

There are two possible directions for future studies. One promising avenue of future work is to make this framework more sophisticated in some respects, such as using covariates of some demographic information for both sides and evaluation of the dependence on prior evaluation. Another direction is additional analyses on the relationship between estimated candidate potential and job performance of the interviewees who are hired. Incorporating this framework into daily operation will make the human resource field even smarter and help decision-makers produce more reliable decisions during hiring process.

\bibliography{sample-base}

\end{document}